\documentclass[amsmath,amssymb,prl,showpacs, twocolumn]{revtex4}

\usepackage{graphicx}
\usepackage{times}

\begin{document}
\title{Laser spot position dependent photothermal mode cooling of a micro-cantilever}

\author{Hao Fu$^{1,2}$}

\author{Cunding Liu$^{1,2}$}

\author{Yong Liu$^{3}$}

\author{Jiaru Chu$^{3}$}

\author{Gengyu Cao$^{1}$*}

\affiliation{$^{1}$State Key Laboratory of Magnetic Resonance and Atomic and
Molecular Physics, Wuhan Institution of Physics and Mathematics, Chinese
Academy of Sciences, Wuhan 430071, P. R. China\\
 $^{2}$Graduate University of the Chinese Academy of Sciences, Beijing
100049, P. R. China\\
 $^{3}$Department of Precision Machinery and Precision Instrumentation,
University of Science and Technology of China, Hefei 230027, P. R.
China}

\date{\today}

\begin{abstract}
We explore the laser spot position (LSP) dependent photothermal mode
cooling of a micro-cantilever in a Fabry-P\'{e}rot (FP) cavity. Depending
on the LSP along the lever, photothermal coupling to the first two
mechanical modes can be either parallel or anti-parallel. This LSP
dependent behavior is analyzed theoretically by a simple model, which
is in quantitatively agreement with our experimen
tal observation.
From simulation, the parallel and anti-parallel coupling region is
identified along the lever. We conclude that a more efficient mode
cooling may be achieved in the parallel coupling region.
\end{abstract}
\pacs{05.40.Jc,42.50.Wk,85.85.+j,42.65.Sf}

\maketitle

Micro-resonator has become an ideal candidate for exploring quantum
phenomena, such as the Heisenberg uncertainty principle and quantum
entanglement, at the boundary between classical and quantum realms
\cite{1,2,3}. Preparing the micro-resonator close to its ground mechanical
state is a crucial step towards observation of such quantum effects
at macroscale \cite{4,5,6}. In addition, cooling operation of the
micro-resonator also attracts enormous interest in diverse areas of
applied science, ranging from ultra-high sensitive measurement \cite{7,8}
to quantum information processing \cite{9}. Recently, cooling schemes
for the micro-resonator have been intensely studied \cite{10,11,12,13},
among which optomechanical schemes are proposed as one of the most
promising strategies to access the macroscopic quantum regime \cite{14,15,16,17,18,19}.
As demonstrated in the pioneer works, the brownian vibrational fluctuations
of the micro-resonator can be quenched significantly by the actively
controlled optical forces of radiation pressure \cite{20}. On the
other hand, implemented through the retarded backaction of optical
forces, passive optical cooling of the optomechanical resonator also
demonstrates the ability of cooling the fundamental mechanical mode
close to ground mechanical state \cite{21,22,23}. Further investigations
on passive optical cooling scheme show that the direction of photothermal
coupling is mechanical mode dependent, which could be either parallel
or anti-parallel for different modes \cite{21,24}. Since anti-parallel
coupling may excite one mode while cooling the other, cooling operation
could be always benefited from the parallel coupling effect. Thus,
it raises an important issue in cooling modes simultaneously. Here
we study the mechanical mode dependence of photothermal mode cooling
at laser spot positions (LSP) along the optomechanical resonator of
compliant micro-cantilever in a low fineness FP cavity. Experiments
at five different LSPs clearly revealed a LSP dependent behavior of
photothermal mode cooling. With the assistance of theoretical analysis,
two types of coupling regions, the parallel coupling region (PCR)
and anti-parallel coupling region (aPCR), are sketched out along the
micro-cantilever according to the relative direction of bolometric
backaction. And we demonstrated that simultaneous cooling of the first
two mechanical modes can be realized at the PCR. It paves the way
for more efficient cooling of an optomechanical resonator to its classical
limit.

\begin{figure}
\centering
\includegraphics[width=\columnwidth]{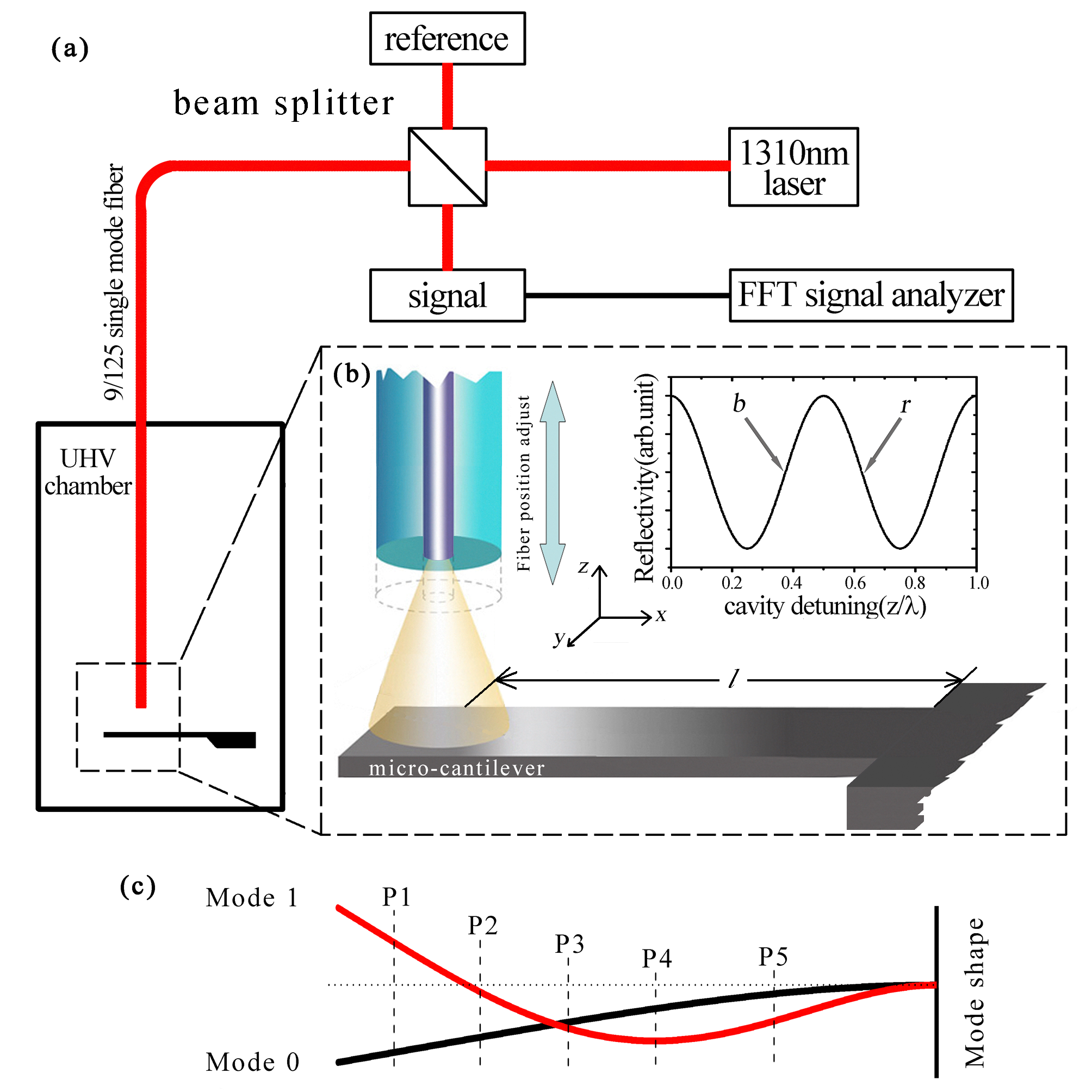}
\caption{(a) Schematics of the experimental setup. (b) The low fineness level-based FP micro-cavity formed by fiber end and micro-cantilever surface. As the inset indicated, the reflectivity is changed periodically with the cavity length. By illuminating the micro-cantilever at position \textit{l}, vibrational resonance of the optomechanical resonator is recorded at the working position marked by \textit{b} and \textit{r} for blue and red cavity detuning accordingly for each laser power. (c) Shapes of mode 0 and mode 1. The photothermal coupling to the first two mechanical modes of micro-lever are investigated at five LSPs denoted by P1 to P5.}\label{Fig-1}
\end{figure}

In our experiment, a single crystal silicon micro-cantilever with
dimension of 480$\mu$m$\times$10$\mu$m$\times$0.8$\mu$m is used \cite{25}. And
gold film of 80nm thickness is thermally evaporated onto the top side
of the lever. The intrinsic resonant frequencies of mode 0 and mode
1 of the micro-cantilever are\textit{ f}$_{0}$= 4443.4Hz and \textit{f}$_{1}$
= 27736.4Hz respectively. In ultra-high vacuum ($8\times10^{-10}$Torr)
and at room temperature, the lever exhibits inherent thermal dissipation
with the damping factors $\Gamma_{0}$=2.9Hz and $\Gamma_{1}$=15.4Hz
for the first two modes. The experimental setup is schematically illustrated
in Fig.\ref{Fig-1}. A laser beam of wavelength $\lambda$=1310nm is supplied
by a semiconductor diode with power programmable from 10$\mu$W to
5mW. After split by a 10dB directional optical coupler (reflection
= 90\%, transmission = 10\%), the laser beam is coupled into the ultra-high
vacuum system by a single mode optical fiber. Aligning the polished
fiber end and the surface of micro-cantilever parallel, a low fineness
level-based FP cavity is formed. A fiber stage piezo capable of accurately
regulating the fiber position in range of $\sim$1$\mu$m
at 300K is employed to switch the working position from the maximum
sensitive interference detection point of red detuning \textit{r} to that of
blue detuning \textit{b} \cite{26}. Laser reflected off the top surface of
micro-cantilever partly couples back to the fiber and interferes with
the laser reflected from the fiber end to produce the micro-cantilever
oscillation signal. The oscillation signal is monitored by a photodetector
and analyzed by an FFT spectrum analyzer (SR760, Stanford Research
System) in the vicinity of the first two resonant frequencies of micro-cantilever.

\begin{figure}
\centering
\includegraphics[width=\columnwidth]{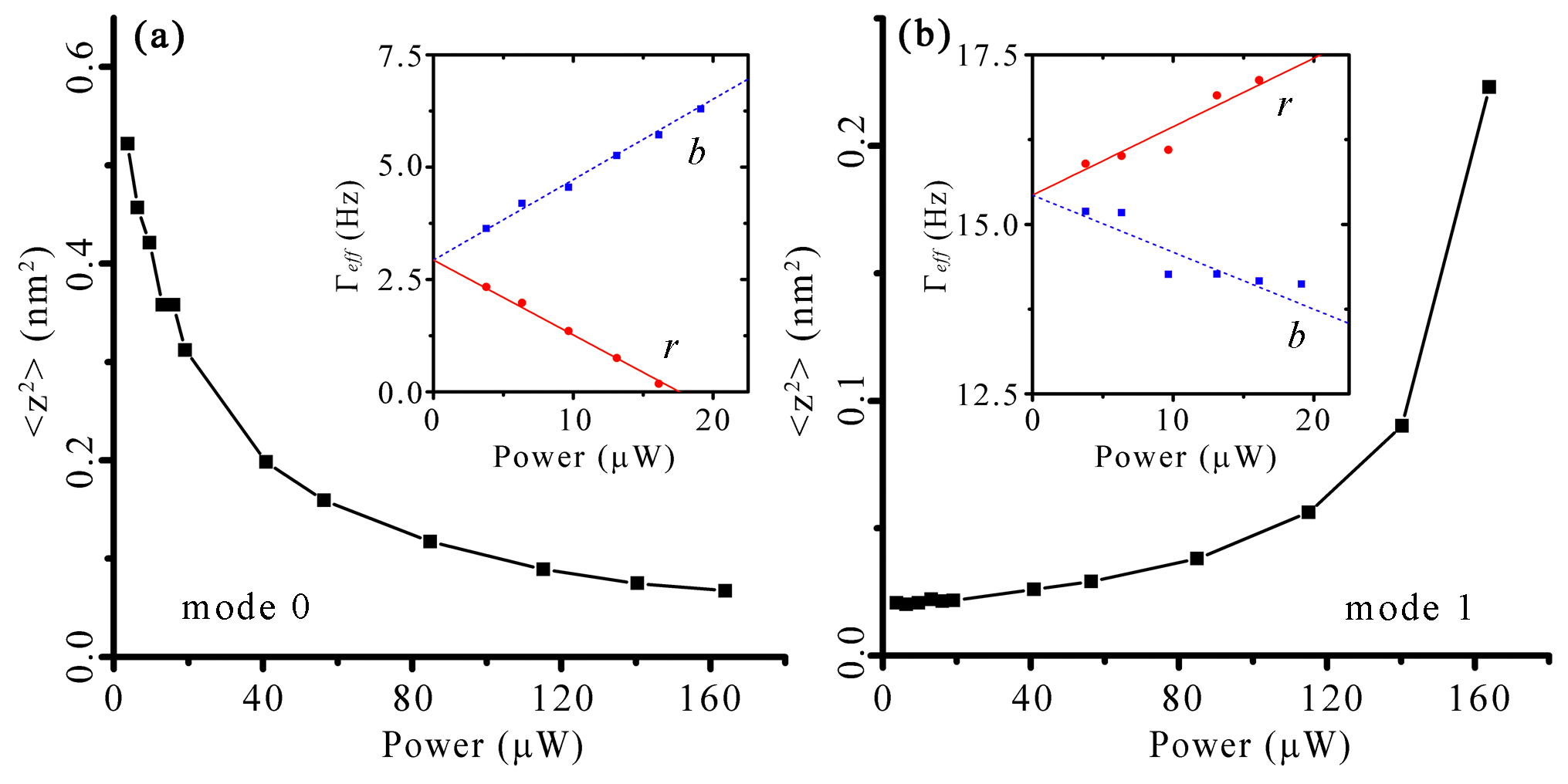}
\caption{Mode dependent bolometric backaction at position P3. Thermal noise amplitude of mode 0 (a) and mode 1 (b) is detected at P3 for the different laser powers. At working point \textit{b}, cooling of mode 0 accompanies with the warming up of mode 1. Insets in graphs (a) and (b) plot the effective damping factors at working points \textit{b} (dash line) and \textit{r} (solid line) against laser power for the two mechanical modes accordingly.}\label{Fig-2}
\end{figure}

The deformable mirror of FP cavity formed by the compliant micro-cantilever
is subject to action of optical forces. It in turn modulates the stored
optical energy and results in the optical forces backaction on the
oscillation of micro-cantilever as a consequence. The mutual modulation
between the laser field inside the cavity and the motion of micro-cantilever
forms the foundation of passive optical cooling. Under the low cavity
fineness condition, only the retarded backaction of photothermal force
(or bolometric backaction) participates in optomechanical cooling
\cite{21}. Other optical forces, such as radiation pressure, react
instantaneously with the oscillation of micro-cantilever, thus modify
only the resonant frequency. Depending on the detuning condition of
micro-cavity, the thermal oscillation of the micro-cantilever can
be either enhanced or suppressed. However, the direction of bolometric
backaction is mechanical mode dependent. As shown in Fig.\ref{Fig-2}, cooling
operation on the fundamental mode (mode 0) always accompanies the
warm-up of mode 1 at position P3 indicated in Fig.\ref{Fig-1}(c). Its thermal
oscillation is enhanced continuously with laser power increasing.
Once crossing the threshold of $\Gamma_{\textit{eff},1}$=0, mode 1 will be
driven into self-sustained oscillation. This result is in consistent
with the observation in the work by G. Jourdan et al \cite{24}.

\begin{figure}
\centering
\includegraphics[width=\columnwidth]{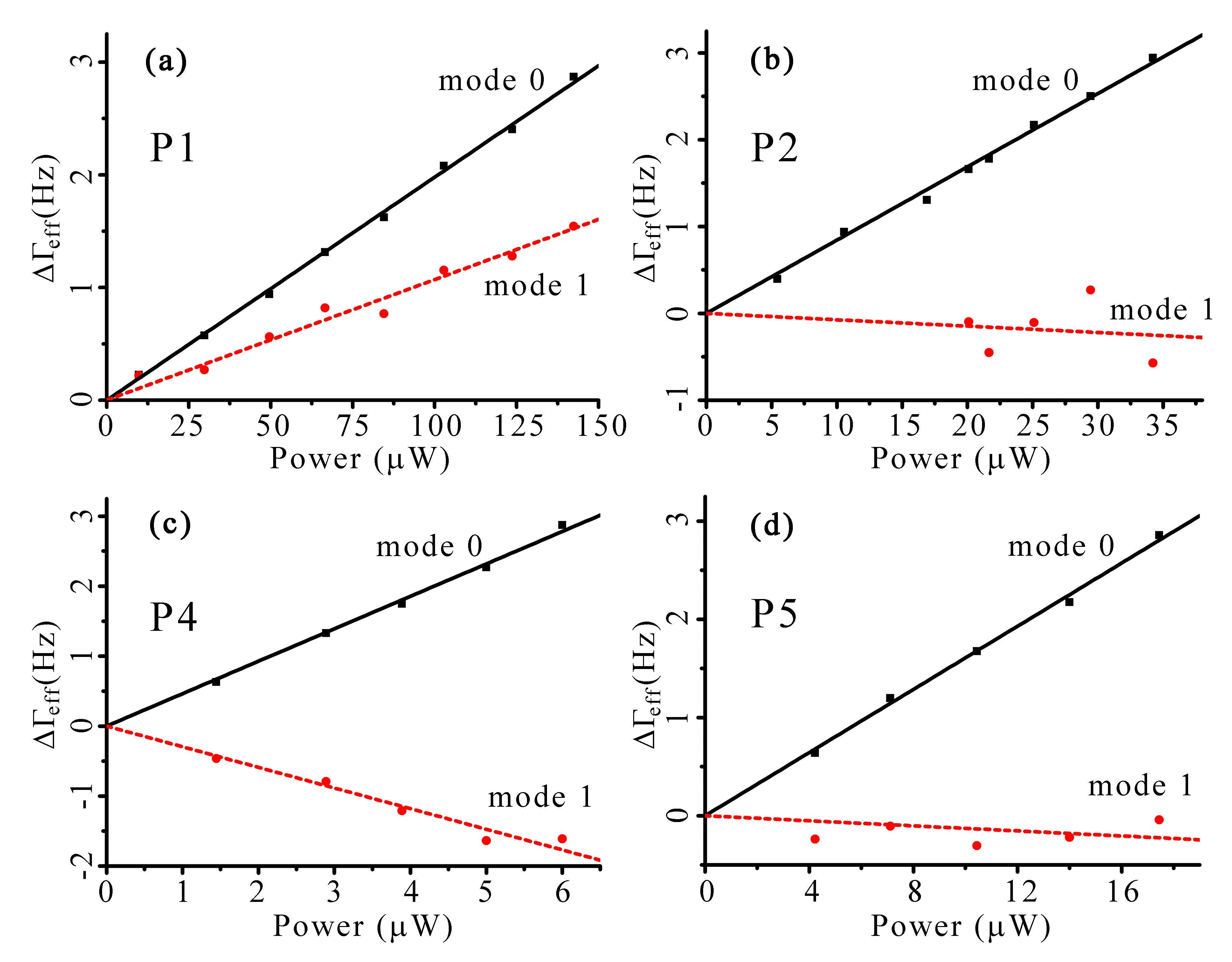}
\caption{Photothermal coupling to mode 0 (solid line) and mode 1 (dash line) at different LSPs. Graph (a) to (d) show the experimental results of the first two modes at P1, P2, P4 and P5 accordingly. During measurements, the cavity length is about 35$\mu$m.}\label{Fig-3}
\end{figure}

This mode dependent photothermal coupling is investigated further
at other LSPs denoted in Fig.\ref{Fig-1}(c). The photothermal damping for mode
n can be represented by $\Delta\Gamma_{\textit{eff},n}=(\Gamma_{\textit{eff},n(b)}+\Gamma_{\textit{eff},n(r)})/2$,
where $\Gamma$$_{\textit{eff},n(b)}$ and $\Gamma$$_{\textit{eff},n(r)}$ are the
effective damping of micro-cantilever at working position \textit{b} and \textit{r}
respectively. In the case of $\Delta\Gamma_{\textit{eff},n}<0$, mode \textit{n} of
the micro-cantilever is enhanced at blue cavity detuning while cooled
at red cavity detuning, and vice versa for $\Delta\Gamma_{\textit{eff},n}>0$.
For mode 0, $\Delta\Gamma_{\textit{eff}}$ is always a positive value, whereas
its value changes with illuminating point for mode 1. As showed in
Fig.\ref{Fig-3}, it indicates a LSP dependent behavior in photothermal mode
cooling. At position P1, the bolometric backaction is exerted on the
two modes in the same direction. Depending on the detuning of cavity,
it provides a possibility of cooling the two modes simultaneously.
Although the bolometric backaction on mode 1 maintains its direction
at those LSPs beyond the vibration node, its relative strength is
varied evidently. The accompanied enhancement of mode 1 at position
P2 and P5 is so weak that its influences become apparent only in the
case of strong photothermal coupling. However, at position P3 and
P4, mode 1 is excited obviously during cooling operation on mode 0.

In order to understand such LSP dependent behavior, we develop a simple
model for the photothermal coupling. As long as the displacement of
micro-cantilever \textit{$u(x,t)$} from its equilibrium position
is the summation of all mechanical modes, denoted by \textit{$u(x,t)=\Sigma\phi_{i}(x)a_{i}(t)$}
where $\phi_{i}(x)$ is the shape of mode \textit{i}, the motion equation
for mode \textit{n} of micro-cantilever driven by both thermal force
\textit{F}$_{th}$ and bolometric force \textit{F}$_{bol}$ can be
expressed as
\begin{equation}
m\ddot{a}_{n}+m\Gamma_{n}\dot{a}_{n}+K_{n}a_{n}=F_{th}-\int^{L}_{0}E\frac{\partial^{2}\phi_{n}}{\partial x^{2}}I_{l}dx
\label{eq:1}
\end{equation}
where \textit{m}, \textit{E}, $\Gamma_{n}$ and \textit{K}$_{n}$
are the effective mass, Young's modulus, damping factor and spring
constant of mode \textit{n} respectively \cite{24}. The integration
term includes the photothermal stress generated all along the length
\textit{L} of the micro-cantilever. A detailed understanding of the
moment of inertia \textit{$I_{l}=\int z\alpha(z)\Delta T_{l}(x,y,z,t)dydz$}
is extremely difficult, because the dynamic temperature field $\Delta T_{l}(x,y,z,t)$
inside the lever body is intricate, which depends on the optical and
thermal properties of lever materials, structural geometry, environment
temperature, laser wavelength and so on. For our gold film coated
micro-cantilever, the photothermal force originates primarily from
the different thermal expansion coefficients between gold film and
silicon material of the lever. The small temperature variation through
the thickness of micro-cantilever is therefore negligible \cite{27}.
Providing the small width of the micro-cantilever as comparing with
the laser spot profile, the temperature distribution can be reduced
to a one dimension field for this composite lever structure. In the
limit of small oscillation amplitude, the oscillation modulated laser
intensity \textit{I}(\textit{z}) is approximated by the Taylor expansion
around equilibrium position \textit{z}$_{0}$: $I(z)\thickapprox I(z_{0})+(z-z_{0})\nabla I(z_{0})$.
Thus the laser induced dynamic temperature field along the micro-cantilever
can be described as:
\begin{equation}
\Delta T_{l}(x,t)=AD_{l}(x)\nabla I(z_{0})\int_{0}^{t}h(t-t')\dot{u}(l,t')dt'\label{eq:2}
\end{equation}
\textit{A} is the absorptivity of laser power. The retarded backaction
of photothermal force is described by the response function $h(t)=1-exp(-t/\tau)$,
where $\tau$ is referred to as the delay time constant of bolometric
backaction \cite{21}. This expression generally takes the contribution
of all mechanical modes into account.

\begin{figure}
\centering
\includegraphics[width=\columnwidth]{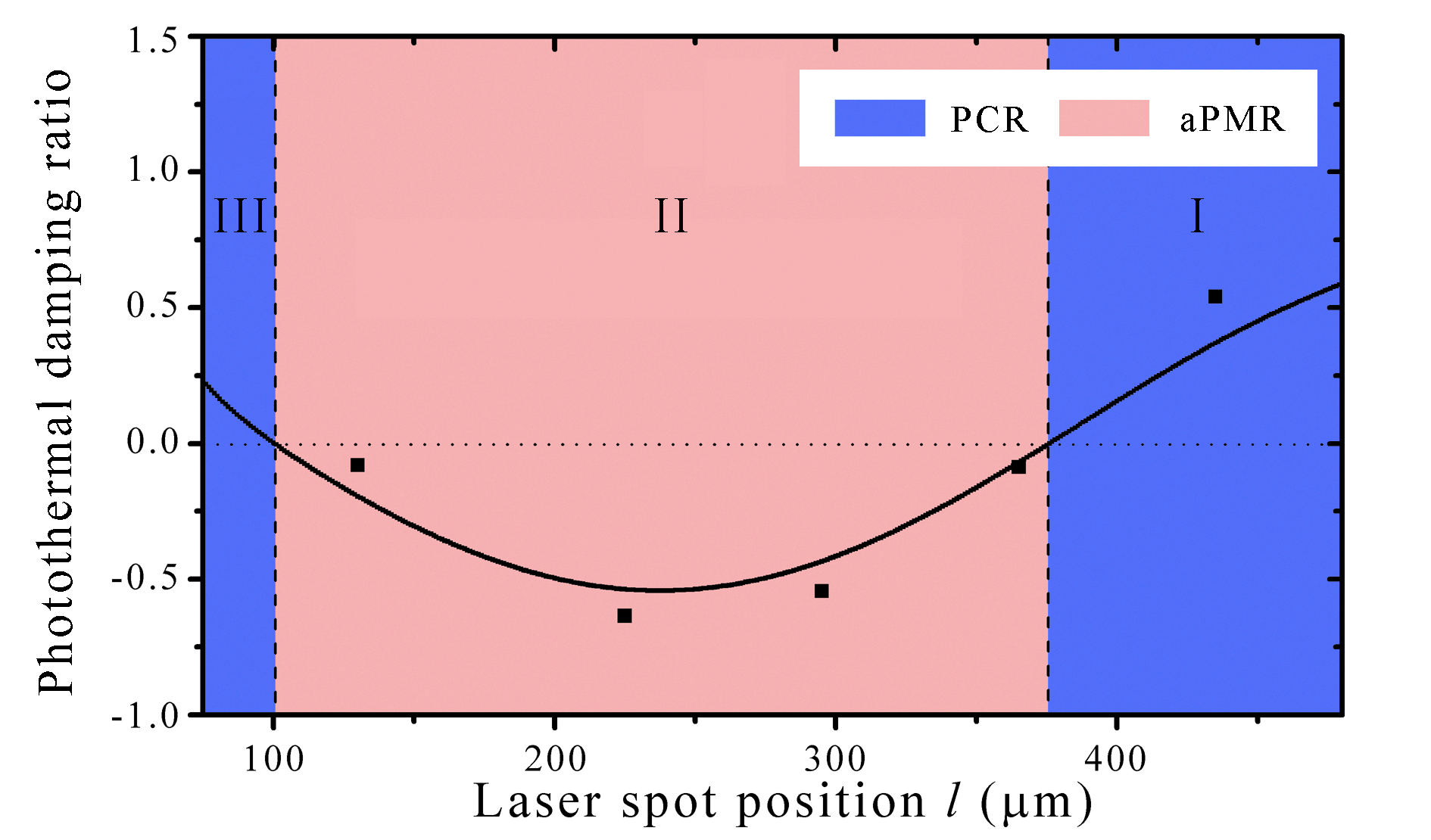}
\caption{Comparison of the experimental measurements with the theoretical simulation. Photothermal damping ratio of mode 1 obtained from experimental data (dots) at the five LSPs indicated in Fig.\ref{Fig-1}(c) are plotted together with theoretical simulation (solid line). Two types of coupling region, the PCR and aPCR, are defined by dash lines.}\label{Fig-4}
\end{figure}

Since the photothermal coupling to higher order mechanical modes is
attenuated by the low pass filter behavior of bolometric backaction,
in this letter, we consider the contribution of the first two modes
only. Combining Eq.\ref{eq:1} and Eq.\ref{eq:2}, the Fourier transform
yields:
\begin{widetext}
\begin{equation}
m\left[\begin{array}{cc}
\omega_{1}^{2}-\omega^{2}+i\omega\Gamma_{1}+\frac{\gamma G_{1}(l)\phi_{1}(l)}{m(1+i\omega\tau)} & \frac{\gamma G_{1}(l)\phi_{2}(l)}{m(1+i\omega\tau)}\\
\frac{\gamma G_{2}(l)\phi_{1}(l)}{m(1+i\omega\tau)} & \omega_{2}^{2}-\omega^{2}+i\omega\Gamma_{2}+\frac{\gamma G_{2}(l)\phi_{2}(l)}{m(1+i\omega\tau)}
\end{array}\right]\left[\begin{array}{c}
a_{0,\omega}\\
a_{1,\omega}
\end{array}\right]=\left[\begin{array}{c}
F_{th,\omega}\\
F_{th,\omega}
\end{array}\right]\label{eq:3}
\end{equation}
\end{widetext}
where $\gamma=AE\nabla I(z_{0})\int z\alpha(z)dydz$ is cavity detuning
condition dependent. In particular, when working position inside the
micro-cavity changes from blue detuning point \textit{b} to the adjacent red
detuning point \textit{r}, the slope of interference $\nabla I(z_{0})$ is reversed
and hence $\gamma$ changes its sign. The mechanical mode dependent
behavior of bolometric backaction is introduced by the function $G_{n}(l)=\int(\partial^{2}\phi_{n}/\partial x^{2})D_{l}(x)dx$.
Since the temperature profile $D_{l}(x)=D_{p}exp(-\left|x-l\right|/r)$
is concentrated around the LSP \cite{28}, it functions to window
the contribution of mode shape in the vicinity of the laser illumination
point only, which decides the LSP dependent nature of photothermal
mode cooling. As Eq.\ref{eq:3} describes, photothermal coupling involves
both self-coupling and inter-modes coupling for each mechanical mode,
which are represented by the diagonal terms and non-diagonal terms
of the matrix, respectively. The contribution of inter-modes coupling
is typically overwhelmed by the self-coupling. Thus, the responses
of the micro-cantilever are dominated by the self-coupling terms,
with the resonant frequency and damping factor modified by the real
and imaginary part of bolometric backaction accordingly. Providing
the relatively small frequency shift, the relative photothermal cooling
effect of mode \textit{n} with respect to fundamental mode can be
represented by the photothermal damping ratio $\beta_{n}$, which
is defined as:
\begin{equation}
\beta_{n}=\frac{d\Delta\Gamma_{\textit{eff},n}}{d\Delta\Gamma_{\textit{eff},0}}=\frac{G_{n}(l)\phi_{n}(l)(1+\omega_{0}^{2}\tau^{2})}{G_{0}(l)\phi_{0}(l)(1+\omega_{n}^{2}\tau^{2})}\label{eq:4}
\end{equation}
Either the bolometric backaction acting in opposite directions on
the two modes simultaneously or not depends completely on the sign
of $\beta_{n}$. Simulation shows that three functional regions of
two types, the PCR and aPCR, are divided for mode 1 along the micro-cantilever
by two nodes, as illustrated in Fig.\ref{Fig-4}. In region I, bolometric backaction
applying onto the two modes of the optomechanical resonator in the
same direction results in a parallel coupling. Crossing the vibration
node of mode 1 such that the sign of $\phi_{1}(l)$ is changed, reversing
of the vibration phase results in altering the direction of bolometric
backaction. Operating in the aPCR of region II, damping of one mode
always accompanies with the enhancement of the other at the same time.
It is not surprising that locally laser heating imposes the second
node in the vicinity of the position $\partial^{2}\phi_{1}/\partial x^{2}=0$,
which imposes the other boundary of aPCR. Beyond this node, in region
III, the mode dependent coupling function $G_{1}(l)$ changes its
sign, whereas mode function $\phi_{1}(l)$ does not. It leads to
the second PCR along the micro-cantilever. By positioning the laser
beam into the PCR, simultaneous cooling of the two modes is achieved
at blue cavity detuning condition. The delay
time constant $\tau$, which sets the cutoff frequency for the bolometric
backaction, is calculated to be 2.5ms for our micro-cantilever at
room temperature \cite{21}. To obtain a more efficient cooling, the
dimension of the micro-cantilever should be designed according to
experiment temperature to satisfy the optimal cooling condition of
$\omega\tau=1$.

It is worth noted here that even higher order mechanical modes of micro-cantilever may become significant in strong photothermal coupling condition. The photothermal cooling involving higher order mechanical
modes can also be analyzed in the same principle. For higher modes,
we predict theoretically that the micro-cantilever could still be
divided into three functional regions, with both boundaries of aPCR
expanding towards two ends of micro-cantilever continuously as the
order of mechanical modes increases as a result. However, the details
inside aPCR become increasingly complicated for higher order modes,
which introduce more nodes within aPCR and emphasize the necessity
of a more accurate description of the temperature field distribution.

In conclusion, by positioning the laser beam onto different points
along the FP optomechanical resonator of the gold film coated micro-cantilever,
we have investigated the LSP dependent behavior of photothermal mode
cooling. Experiments show that not only the direction but also the
efficiency of photothermal mode cooling is LSP dependent. According
to the direction of bolometric backaction, we divide the micro-cantilever
into three coupling region of two types, the PCR and aPCR. Simultaneous
cooling of the first two modes can be achieved when laser beam is
pointed onto the PCR of micro-cantilever. After satisfying the optimal
cooling condition such that the residual thermal effect is minimized,
photothermal cooling limitation may ultimately reach by illuminating
the micro-cantilever at the cross section of the PCRs of the first
few mechanical modes.

This work was supported by the Grand Project of Instrumentation and
Equipments for Scientific Research of the Chinese Academy of Sciences
under contract YZ0637. We gratefully acknowledge Y. Miyatake in Unisoku
Scientific Instruments for technical supports.

\end{document}